\documentclass[preprint,showpacs]{revtex4}

\usepackage{graphicx}

\newcommand{\be}[1]{\begin{equation}\label{#1}}
\newcommand{\ee}{\end{equation}}

\begin{document}

\title{Rational Expectations, psychology and inductive learning via
  moving thresholds}

\author{H. Lamba} \affiliation{Department of Mathematical Sciences,
George Mason University, 4400 University Drive,
Fairfax, VA 22030 USA}

\author{T. Seaman} \affiliation{School of Computational Sciences,
George Mason University, 4400 University Drive,
Fairfax, VA 22030 USA}

\date{August 30th 2007}

\begin{abstract}

This work suggests modifications to a previously introduced 
class of heterogeneous agent models 
that allow for the inclusion of different types of agent  motivations   
and behaviours in a unified way.
The agents operate within a highly simplified environment where they
are only able to be long or short one unit of the asset. The price of
the asset is influenced by both an external information stream and the
demand of the agents. The current strategy of each agent is defined by a pair
of moving thresholds straddling the current price. When the price
crosses either of the thresholds for a particular agent, that agent switches position and a new
pair of thresholds is generated. 

Different kinds of threshold motion 
 correspond to different sources of motivation, running the gamut from
 purely rational information-processing, through rational (but often
 undesirable) behaviour
 induced by perverse incentives and moral hazards, to purely
 psychological effects.
As with the previous class of models, the fact that the simplest model
of this kind precisely conforms to the Efficient Market Hypothesis
allows causal relationships to be established between properties at
the agent level and violations of EMH price statistics at the
global level.   

\end{abstract}


\maketitle

\section{\label{intro}Introduction}  

The Efficient Market Hypothesis (EMH) \cite{f70}
still has enormous influence in political and economic theory as well
as in the day-to-day operations of financial markets. This is in spite
of numerous statistical and experimental studies that invalidate both
the underlying assumptions (eg. the rationality of economic
participants) of the EMH and the output of mathematical models that
are derived from them (eg. stock prices that are described by a  geometric Brownian
motion resulting in a Gaussian distribution of the changes in the log-price returns) \cite{ms00,c01}. 
The near-universal statistical properties of real markets, most of
which deviate
from those predicted by the EMH, are often referred to as the
`stylized facts'.

The inappropriate use of EMH models and Gaussian statistics by market
participants can cause great damage, not just to the participants
themselves but to entire economies. 
For example, a
widely-used form of portfolio risk management, called Value-at-Risk,
attempts to estimate the maximum loss that can occur with a specified
(low) probability. Also many so called `quantitative' hedge-funds use
historical correlations between different financial products to
identify and exploit perceived market mispricings. This
extrapolation from the past into the future becomes less justifiable
in an environment where the decreasing probability of large
fluctuations follows a power-law rather than an exponential one
(i.e. where `fat-tails' are present) and significant coupling between
market participants can develop.

To take a very recent example, at the time of writing 
a sudden rise in the delinquency rates of so-called subprime
mortgages in the US has precipitated a credit-crunch in financial
markets around the world. The growth of this industry was stimulated
by a period of low global interest rates, inadequate accounting and
credit-rating standards and 
the development of novel
financial derivatives such as Collateralized Debt Obligations (CDOs)
and Credit Default Swaps (CDSs).

However it is the proximal causes that are of more  interest us
here. The herding phenomenon and the resulting `madness of crowds'
that lies at the heart of most, if not all, financial bubbles was
present in the general public and their increasingly irrational belief
that house prices would continue to rapidly climb. This caused a cycle of
additional house-buying and home-equity withdrawal that fuelled prices
further. However, on the lending side of the equation, there were
significant perverse incentives and moral hazards at
play. Realtors, property appraisers and mortgage brokers all receive
their commissions at the time of the transaction and have no incentive to  
question the quality of the transaction or lower its amount. The loan
originators were similarly able to distance themselves from the
default risk on the loans via the use of CDOs and CDSs. This continued
up the financial food chain, helped by the fact that there was no open 
market for these derivatives and their prices could be easily manipulated  
via creative accounting practices. Finally, when the housing market burst
the liquidity assumption inherent in almost all financial models ---
that any desired transaction can be carried out at the current
price --- failed to hold as buyers of these derivatives vanished from
the marketplace. 

In short, it is hard to see how models
based solely upon the notion of efficient markets can begin to adequately predict or
quantify situations such as the one described above. The construction
of financial models that can incoporate both rational and  irrational 
agent behaviour, as well as the rational-but-perverse consequences of
market defects, is an important undertaking. The simple,
yet plausible, class of models introduced in this paper provides a
possible framework within which the consequences of different EMH violations can be
systematically studied.

\section{\label{sec2}Previous Results}

In \cite{cgls05,cgl06,ls06,cgls07} a class of models using thresholds
was introduced. These are briefly described below but the reader is directed
to the references for further details. The models consisted of two
parts, one defining the  price updates and the other  determining when
agents switch. The price changes were determined by both external factors,
namely a Gaussian uncorrelated information stream, and
changes in the internal supply/demand due to agents switching
positions.

The thresholds are involved in the modeling of the agents themselves. In the simplest case,
whenever an agent switches position, a pair of {\em fixed} price
thresholds  is generated (from some predetermined distribution) that
straddles the current price. Then, when the price crosses either of
the thresholds the agent switches and a new pair of thresholds is
generated. If one interprets the price thresholds as representing the agent's
rational future expectations then the model is, both practically and
philosophically, an EMH model. Differing expectations cause trading to
occur but the lack of any kind of coupling between agents ensures
the `correct' pricing outcome.

A propensity towards herding, either due to subconscious psychology or 
conscious momentum-trading strategies, was then introduced via the
introduction of an additional threshold for each agent. While an agent
is in the minority their herding pressure increases until they switch
to join the majority (unless the majority position changes, or they
switch due to the price thresholds being violated, before this
occurs). The main conclusion to be deduced from these early models was
that, at least within this modeling scenario, herding causes fat-tails
and excess kurtosis in the price return data but not volatility
clustering. However, long-term correlations in the volatility could be
induced by allowing the volatility to depend upon the market sentiment
and/or allowing correlations in the information stream. A
detailed statistical analysis \cite{ls06} showed that all of the
important 
stylized facts could be reproduced and others, such as the asymmetry
between large positive and negative moves, could also be introduced
using asymmetric herding.

\section{\label{sec3}Moving-threshold Models}

The inclusion of additional non-EMH effects, using the above approach,
would require more thresholds. However,a much more elegant
threshold-based approach is to allow the original pair of
price-thresholds to vary with time between switchings. The full
mathematical moving-threshold model is now described and more detailed
economic justifications for the modeling assumptions can be found in
\cite{cgls05,ls06,cgls07}.

The system is incremented in
timesteps of length $h$ and each of $M$ agents can be either short or
long the asset over each time interval.
The position of the $ i^{th} $ investor over the $ n^{th} $ time
interval is represented by
  $s_i(n) = \pm 1 $ ($+1$ long, $-1$ short).
The price of the asset at the end of the
$n^{\rm th}$ time interval is $p(n)$ and  for simplicity the system is
 drift-free so that $p(n)$ corresponds to the return relative to
the risk-free interest rate plus equity-risk premium or
 the expected rate of return.
An important variable is the {\em sentiment} defined as 
the average of the states of all of the $M$ investors
\begin{equation} \sigma (n) = \frac{1}{M} \sum_{i=1}^M s_i(n).\label{sigma} \end{equation}
and $\Delta \sigma (n) = \sigma (n) - \sigma(n-1)$.

The pricing formula is given by 
  \begin{equation}
p(n+1)=p(n) \exp\left(\left(\sqrt{h}\eta(n) - h/2\right)
f(\sigma) + \kappa \Delta\sigma(n)\right)\label{price1}
\end{equation}
where $\kappa > 0$ and $\sqrt{h}\eta(n) \sim {\cal N}(0,h)$ represents the exogenous
information stream. 
Note that when $\kappa = 0$ and $f\equiv 1$ the price follows a
geometric Brownian motion. 

Suppose that at time $n$ the $i^{\rm th}$ investor has just switched and the
current price is $P$. Then a pair of numbers $X_L,X_U >0$ are generated
from some random distribution and the lower and upper thresholds for that agent are
set to be $L_i(n) = P/(1+X_L)$ and $U_i(n)= (1+X_U)P$ respectively.
Defining the evolution of these thresholds now corresponds to defining
a {\em strategy} for the $i^{\rm th}$ investor. Such strategies can be
made arbitrarily complicated and may be
partly rational (and conscious) and  partly irrational (and
subconscious). They can also be constructed to include perverse
incentives or inductive learning strategies as required. 

Three simple examples are of particular note. Firstly, as noted above, the case
where the thresholds are fixed (until the agent switches again)
completely decouples the agents' behaviour and gives EMH pricing. Secondly, we can mimic the
herding effect by causing the thresholds to move together (increasing
$L_i$ and decreasing $U_i$) whenever the $i^{\rm th}$
agent is in the minority. Thus agents in the minority have a higher
tendency to switch into the majority than vice versa. Thirdly, we can suppose that a
simple, unspecified,  perverse incentive is in place, causing
agents to prefer one of the states over the other, say $+1$ over
$-1$. This can be recreated by moving the thresholds closer together
whenever $s_i(n)=-1$. 

The threshold approach should be contrasted with the more common
one taken in the literature of Markovian-type switching between
investment positions or strategies (\cite{lm00} and numerous
others). Here the threshold values $L_i,U_i$ act as `hidden variables'
that are highly-history dependent. Much of human behaviour is, of
course, also non-Markovian with decisions being made over a period of
time, rather than being spontaneously formulated and acted upon.
Finally, we emphasize that the purpose of such simple models is to help gain
insights into EMH violations and their causal relationships to
the observed statistical properties of real markets.

\section{\label{sec4}Numerical results}  

We first select parameters for the model to simulate daily price variations.
The time variable $h$ is defined in terms of the variance of the
external information stream. A daily variance in price returns of 0.6--0.7\%
suggests a value for $h$ of 0.00004. 
The system properties are independent of the number of agents, for
large enough
$M$, and $M=100$ appears to be sufficient.
The simulations are run for 10000 timesteps which corresponds to
approximately 40 years of trading.

The parameter $\kappa$ is a measure of the market depth and the relative
importance of external noise versus changes in internal supply/demand
on the asset price. It is difficult to
estimate a priori, but $\kappa = 0.2$ generates price output that is
correlated with, but noticably distinct from, the EMH price 
defined by $\eta(n)$ alone. 

The initial thresholds $X_L,X_U$ are chosen from the uniform
distribution on the interval $ [0.1,0.3]$, corresponding to price moves
in the range 10--30\%. Simulations have indicated  that the
models are robust to changes in the exact form of the distribution 
and so a uniform distribution was chosen.
Let us consider first the case where the thresholds are fixed. 
If  $f(\sigma) \equiv 1$ and the initial states are sufficiently mixed
so that $\sigma(0) \approx 0$, then
as
explained  in Section~\ref{sec3}, the model conforms to the EMH and
the price is simply 
  \begin{equation}
p(n+1) \approx p(n) \exp\left(\sqrt{h}\eta(n) - h/2
 \right). \label{emh}
\end{equation}

As argued in \cite{cgls05,ls06,b99}, the effect of noise traders 
(who are not explicitly  included) is unlikely
to be constant over time. We posit that their number, and therefore
also their effect, is greater at times of high sentiment, both
positive and negative. This is simulated by increasing the
price-changing effect of the external information stream at such times
by the use of the function $$ f(\sigma) = 1 + 2 |\sigma|$$ in (\ref{price1}).

\begin{figure}
  \includegraphics[width=3in]{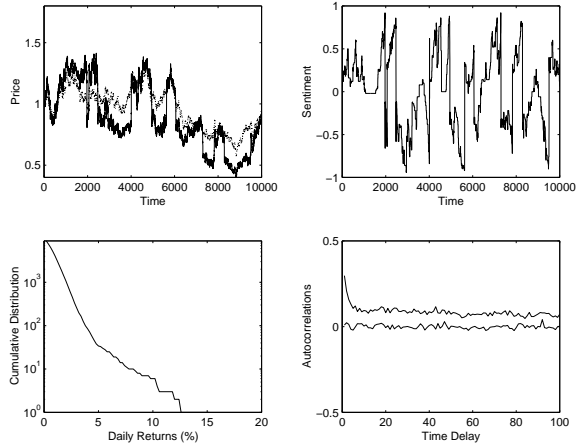}
\caption{\label{fig1} Results of a simulation over 10000
  timesteps including herding.}
\end{figure}

Now let us introduce herding by supposing that for agents in the
minority position
$$ L_i(n+1) = L_i(n) + C_i h |\sigma(n)|, \quad 
 U_i(n+1) = U_i(n) - C_i h |\sigma(n)|. $$
The thresholds are fixed for agents in the majority position. Note
that the change in the position of the thresholds is proportional to
the length of the timestep and the magnitude of the sentiment. The
constant of proportionality $C_i$ is different, but fixed, 
for each agent and chosen from the uniform distribution on
$[0,100]$. This range of parameters corresponds to a herding tendency
that operates over a timescale of several months or longer. The
results of a simulation are shown in Figure~\ref{fig1} and are very
similar to those obtained in \cite{cgls05,ls06} where multiple, fixed
thresholds were used.

The top left plot shows both the output price $p(n)$ (more volatile) and the EMH pricing
obtained from (\ref{emh}) (less volatile). As can be seen there are significant
periods of price mismatching. The top right figure plots the sentiment
against time. Periods of bullish and bearish sentiments lasting
several years can be observed. The bottom left picture shows clear
evidence of a fat-tailed distribution in the price returns and this is
also confirmed by measures of the excess kurtosis which range from
approximately 10--30. Finally, the two curves in the bottom right
figure are the autocorrelation decays for both the price returns and
their absolute values (the volatility). There is no correlation
observable in the price returns, even for lags of a single day, while
the volatility autocorelation decays slowly over several months ---
evidence of volatility clustering. Further numerical testing revealed
very similar conclusions to those drawn from the previous
fixed-threshold models. These were that, in these models,  herding causes fat tails but
not volatility clustering since the decay in the volatility
autocorrelation vanishes after just a few days when $f(\sigma) =
1$. Also, measurements of power-law exponents similar to those carried
out in \cite{ls06} provided estimates close to those observed in
analyses of price data from real markets.

\begin{figure}
  \includegraphics[width=3in]{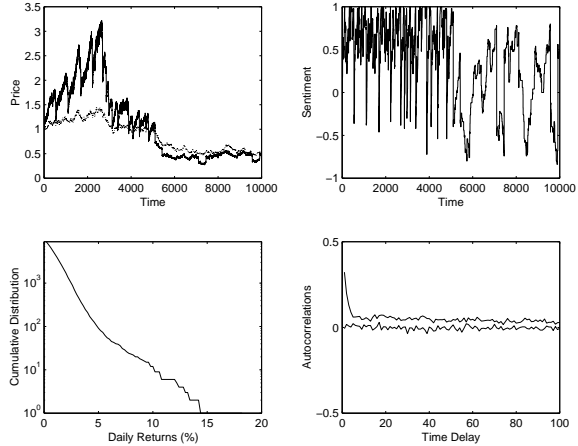}
\caption{\label{fig2} Results of a simulation over 10000
  timesteps including both herding and a perverse incentive. The
  incentive is removed after 5000 timesteps.}
\end{figure}

We now suppose that, in addition to the herding and
sentiment-dependent volatility, a perverse incentive is in place that
makes it advantageous to be in the state $+1$. This is achieved by
including the following extra rule when an agent is in the state
$s_i(n) = -1$
$$ L_i(n+1) = L_i(n) + R h, \quad 
 U_i(n+1) = U_i(n) - R h. $$
The value $R=100$ was chosen to make the incentive  the same order of
magnitude as the herding propensity. The incentive is then switched
off after 5000 timesteps. Results for a typical simulation are
shown  in Figure~\ref{fig2}.
 
The effect of the incentive can be seen in both the price and sentiment
variables, where both the average value and the volatility of the variables
are increased. This simulation merely serves as a prototype for the kinds of
studies that can be performed and  more systematic investigations are
clearly required in order to draw reliable inferences. These will be
reported elsewhere.

\section{\label{sec5}Conclusions}  

In this paper an agent's {\em strategy} (in the loosest sense
of the word) is defined to be a combination of all the rational,
inductive learning, psychological
and rational-but-perverse factors influencing their decision to switch
positions or stay put. We have shown that such strategies can be
represented by moving thresholds and can be as simple or as
complicated as desired. Indeed the most serious modeling restrictions
in the caricature systems simulated above are not in the assumptions
behind the moving threshold approach, but in the assumptions made upon the
market itself. Giving agents the option to leave the market, a choice
of assets, and allowing their strategies to depend upon their current
wealth are all obvious first directions in which the models could be
extended. However, it seems unlikely that the causal relationships
between strategies, as defined above, and global market properties
will be understood in complex systems until they are understood in
simpler ones.


\end{document}